\documentclass[a4paper,fleqn,usenatbib]{mnras}

\usepackage{graphicx}	% Including figure files
\usepackage{amsmath}	% Advanced maths commands
\usepackage{amssymb}	% Extra maths symbols
\usepackage{pdflscape}
\def\msun{\hbox{M$_\odot$}}

\title[The open cluster Haffner\,9]{Towards a comprehensive knowledge of the open 
cluster Haffner\,9}

\author[A.E. Piatti]{
Andr\'es E. Piatti$^{1,2}$\thanks{E-mail: andres@oac.unc.edu.ar}\\
$^{1}$Observatorio Astron\'omico, Universidad Nacional de C\'ordoba, Laprida 854, 5000, 
C\'ordoba, Argentina\\
$^{2}$Consejo Nacional de Investigaciones Cient\'{\i}ficas y T\'ecnicas, Av. Rivadavia 1917, 
C1033AAJ, Buenos Aires, Argentina\\
}

\date{Accepted XXX. Received YYY; in original form ZZZ}

\pubyear{2016}

\begin{document}
\label{firstpage}
\pagerange{\pageref{firstpage}--\pageref{lastpage}}
\maketitle

% Abstract of the paper
\begin{abstract}
We turn our attention to Haffner\,9, a Milky Way open cluster
whose previous fundamental parameter estimates are far from being in agreement.
In order to provide with accurate estimates we present high-quality Washington 
$CT_1$
and Johnson $BVI$
photometry of the cluster field. We put particular care in statistically clean
the colour-magnitude diagrams (CMDs) from field star contamination, which was found
a common source in previous works for the discordant fundamental parameter estimates. 
The resulting cluster CMD fiducial features were confirmed from a proper motion 
membership analysis. Haffner\,9 is a moderately young object (age $\sim$ 350 Myr),
placed in the Perseus arm -at a heliocentric distance of $\sim$ 3.2 kpc-, with
a lower limit for its present mass of $\sim$ 160 $\msun$ and of nearly metal solar content.
The combination of the cluster structural and fundamental parameters suggest
that it is in an advanced stage of internal dynamical evolution, possibly in the phase 
typical of those with mass segregation in their core regions. However, the cluster 
still keeps its mass function close to that of the Salpeter's law.
\end{abstract}

% Select between one and six entries from the list of approved keywords.
% Don't make up new ones.

%%%%%%%%%%%%%%%%%%%%%%%%%%%%%%%%%%%%%%%%%%%%%%%%%%

%%%%%%%%%%%%%%%%% BODY OF PAPER %%%%%%%%%%%%%%%%%%

\section{Introduction}

Different statistical procedures have been proposed with an acceptable success,
in order to avoid as much as possible the field contamination in cluster 
colour-magnitude diagrams (CMDs) analysis \citep[e.g.][]{bb07,pb07,maiaetal2010}
The developed statistical methods basically involve: 
(i) dividing the full range 
of magnitude and colour of a given CMD into a grid whose cells have axes along the magnitude and colour directions, (ii) computing the expected number-density of field stars in each cell based on the number of comparison
field stars with magnitude and colour compatible with those of the cell, and (iii) subtracting randomly the expected 
number of field stars from each cell. Although the methods reapply the cleaning 
procedure using different cell sizes in the CMDs, they are fixed each time, i.e., they do not vary across the CMDs. 

From our experience in cleaning the field star contamination in the cluster CMDs, we have identified some situations
which still need our attention. It frequently happens that some parts of the CMDs are more populated than others, so 
that fixing the size of the cells in the CMDs becomes a difficult task. Small cells do not usually carry out a 
satisfactory job in CMD regions with a scarce number of fields stars, while big cells fail in populous CMD regions. 
Thus, relatively
bright field red giants with small photometric errors could not be subtracted and, consequently, the cluster
CMD could show spurious red giant features.  A compromise 
between minimizing the residuals left after the subtractions of field stars from the cluster CMDs and maximizing the 
cleaning of field stars is always desiderable. 

In this paper, we present a comprehensive multi-band photometric analysis of 
Haffner\,9 from $BVI$ and Washington $CT_1$ photometry. The cluster has been
previously studied from 2MASS and $BVI$ photometry. However, a relatively lower
photometric accuracy and shallower limited magnitude, in combination
with a not much pinpointed treatment of field decontamination led those studies
to discordant results. In Section 2 we describe the collection
and reduction of the available photometric data and their thorough treatment in order to build a extensive and reliable data set. The cluster structural and fundamental parameters are derived
from star counts and colour-magnitude and colour-colour diagrams, respectively, as described in Section 3.
The analysis of the results of the different astrophysical parameters obtained is carried out
in Section 4, where implications about the stage of its dynamical evolution are suggested.
Finally, Section 5 summarizes the main conclusion of this work.

\section{Data collection and reduction}

We used CCD images\footnote{The images are made available to the public through
http://www.noao.edu/sdm/archives.php, SMARTS Consortium, DDT, PI: Clari\'a.} obtained on the nights of December 18$^{\rm th}$ 
and 19$^{\rm th}$, 2004 with a 2048$\times$2048 pixel Tektronix CCD attached to the 0.9 m telescope 
(scale 0.396 arcsec/pixel) at Cerro Tololo Inter-American Observatory (CTIO, Chile). 
Its  field of view is 13.6$\times$13.6 arcmin$^2$. We used the Washington $C$ 
\citep{c76} and Kron-Cousins $R$ filters. The latter has a
significantly higher throughput as compared with the standard Washington $T_1$ filter so that $R$ magnitudes can be accurately
transformed to yield $T_1$ magnitudes \citet{g96}. We used a series 
of bias, dome and sky flat-field exposures per filter to calibrate the CCD instrumental signature. 
We also utilised images for the Small Magellanic 
Cloud (SMC) cluster Lindsay\,106, which was previously observed at La Silla (ESO, Chile) 
with the $C$ and $T_{1}$  filters \citep{petal07b}. Lindsay\,106 was used here only 
as control cluster, i.e., to verify the quality of the present CTIO photometry. Table 1 shows the log 
of the observations with filters, exposure times,  airmasses and seeing estimates. 
A large 
number (typically 20) of standard stars from the list of \citet{g96} was also observed 
on each night. They cover wide colour and airmass ranges, so that we could calibrate properly the 
program stars observed on these nights.

\begin{table*}
\caption{Observations log of the open cluster Haffner\,9 and the control field Lindsay\,106.}
\label{tab:table1}
\begin{tabular}{@{}lccccccccc}\hline
Cluster$^a$  &R.A.      &Dec.     &{\it l} &b       & date & filter & exposure & airmass & seeing\\
         &(h m s)   &($\degr$ $\arcmin$ $\arcsec$)&(\degr)&(\degr)&  & &  (sec)  &  & ($\arcsec$)\\
\hline
Lindsay\,106, ESO\,29-SC44  & 1 30 38 &-76 03 16 & 299.82 & -40.84 & Dec. 19 & $C$ & 2400 & 1.47 & 1.4 \\
                      &         &          &        &        &         & $R$ &  900 & 1.45 & 1.3 \\
Haffner\,9            & 7 24 42 &-17 00 10 & 231.80 & -0.59  & Dec. 18 & $C$ &  300 & 1.05 & 1.3\\
                      &         &          &        &        &         & $C$ &  300 & 1.05 & 1.3\\
                      &         &          &        &        &         & $R$ &   10 & 1.06 & 1.1\\
                      &         &          &        &        &         & $R$ &   10 & 1.06 & 1.1\\
                      &         &          &        &        &         & $R$ &   30 & 1.07 & 1.2\\
                      &         &          &        &        &         & $R$ &   30 & 1.07 & 1.2\\

\hline
\end{tabular}

\noindent $^a$ Cluster identifications are from \citet[][L]{l58} and \citet[][ESO]{lauberts1982}.
\end{table*}

The stellar photometry was performed using the star finding and
point spread function (PSF) fitting routines in the {\sc DAOPHOT/ALLSTAR}
suite of programs \citep{setal90}. Radially varying aperture corrections were applied to take out the 
effects of PSF variations across
the field of view, although a quadratically varying PSF was employed. The resultant instrumental 
magnitudes 
were standardized using the equations:

\begin{equation}
\begin{split}
c = (3.679 \pm 0.021) + T_1 + (C-T_1) + (0.294 \pm 0.014)\times X_C\\
-(0.085 \pm 0.005) \times (C-T_1),
\end{split}
\end{equation}

\begin{equation}
\begin{split}
r = (3.206 \pm 0.021) + T_1 + (0.115 \pm 0.014)\times X_{R} \\
-(0.014 \pm 0.004)\times (C-T_1),
\end{split}
\end{equation}

\noindent where $X$ represents the effective airmass. Capital and lowercase letters stand for 
standard and instrumental magnitudes, respectively. The coefficients were derived through the 
{\sc IRAF\footnote{IRAF is distributed by the National Optical Astronomy 
Observatories, which is operated by the Association of Universities for Research in Astronomy, 
Inc., under contract with the National Science Foundation}} routine {\sc FITPARAM}.
The root mean square (rms) deviations of the fitted values from the
fits to the standards were 0.021 for $c$ and 0.014 for $r$, which 
indicates that the nights were photometric. We combined all the independent measurements using the 
stand-alone {\sc DAOMATCH} and {\sc DAOMASTER}
programmes,
kindly provided by Peter Stetson. The final information gathered for each cluster consists of a 
running number per star, the $x$ and $y$ coordinates, the measured $T_1$ magnitudes, $C-T_1$ colours, 
and the observational errors $\sigma(T_1)$ and $\sigma(C-T_1)$. 
In the case of Haffner\,9,
for which we combined respectively four different photometric tables, we also included the number of measures 
performed in $T_1$ and $C-T_1$. Table~\ref{tab:table2}  gives this information, where only a portion 
of it is shown for guidance regarding its form and content. The whole content 
of Table~\ref{tab:table2}, as well as that for Lindsay\,106 (Table 3), is 
available in the online version of the journal.

\begin{figure}
	\includegraphics[width=\columnwidth]{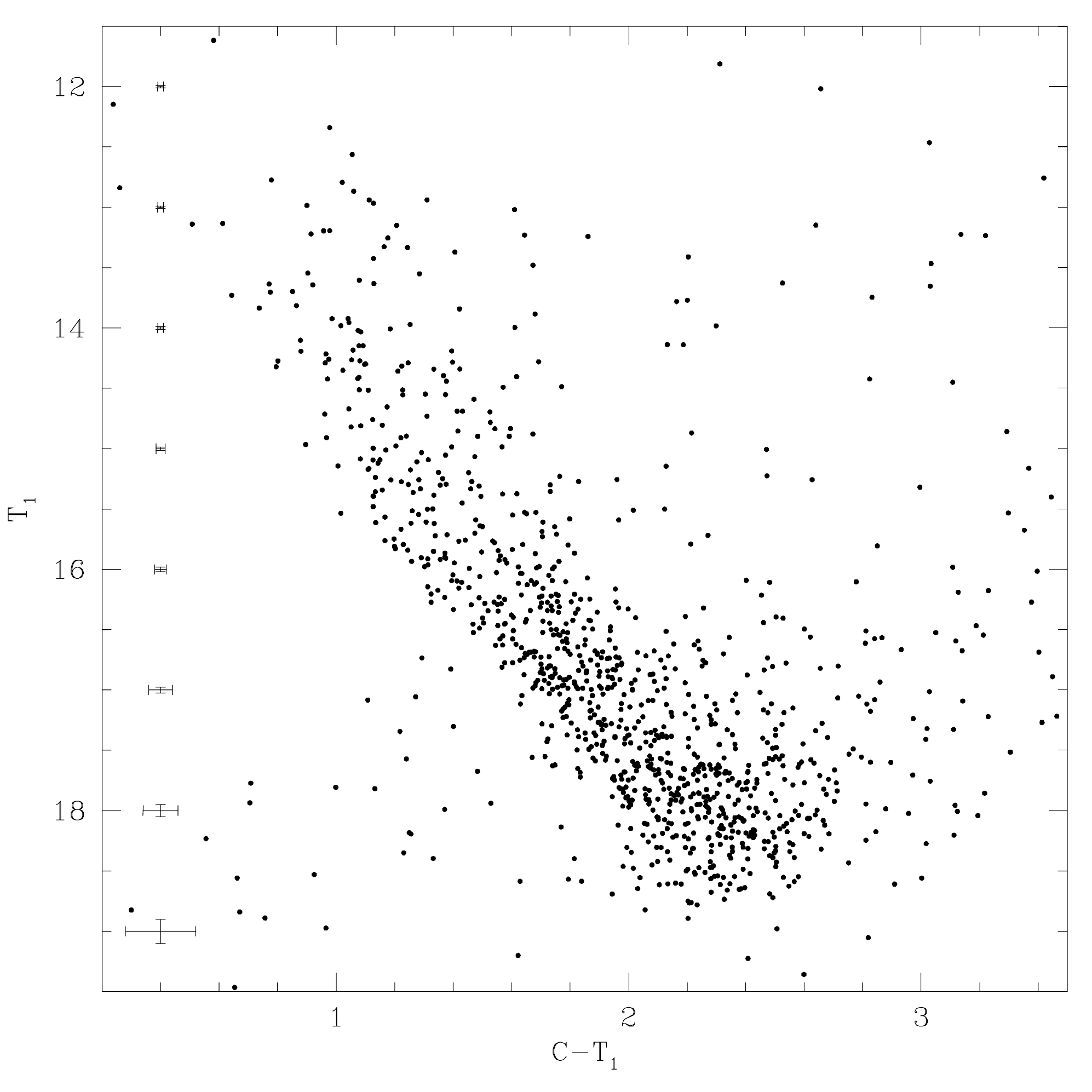}
    \caption{Observed CMD for all the stars measured in the field of haffner\,9. Errorbars
at the left-hand margin represent the photometric uncertainties given by DAOPHOT.}
    \label{fig:fig1}
\end{figure}

\begin{figure}
	\includegraphics[width=\columnwidth]{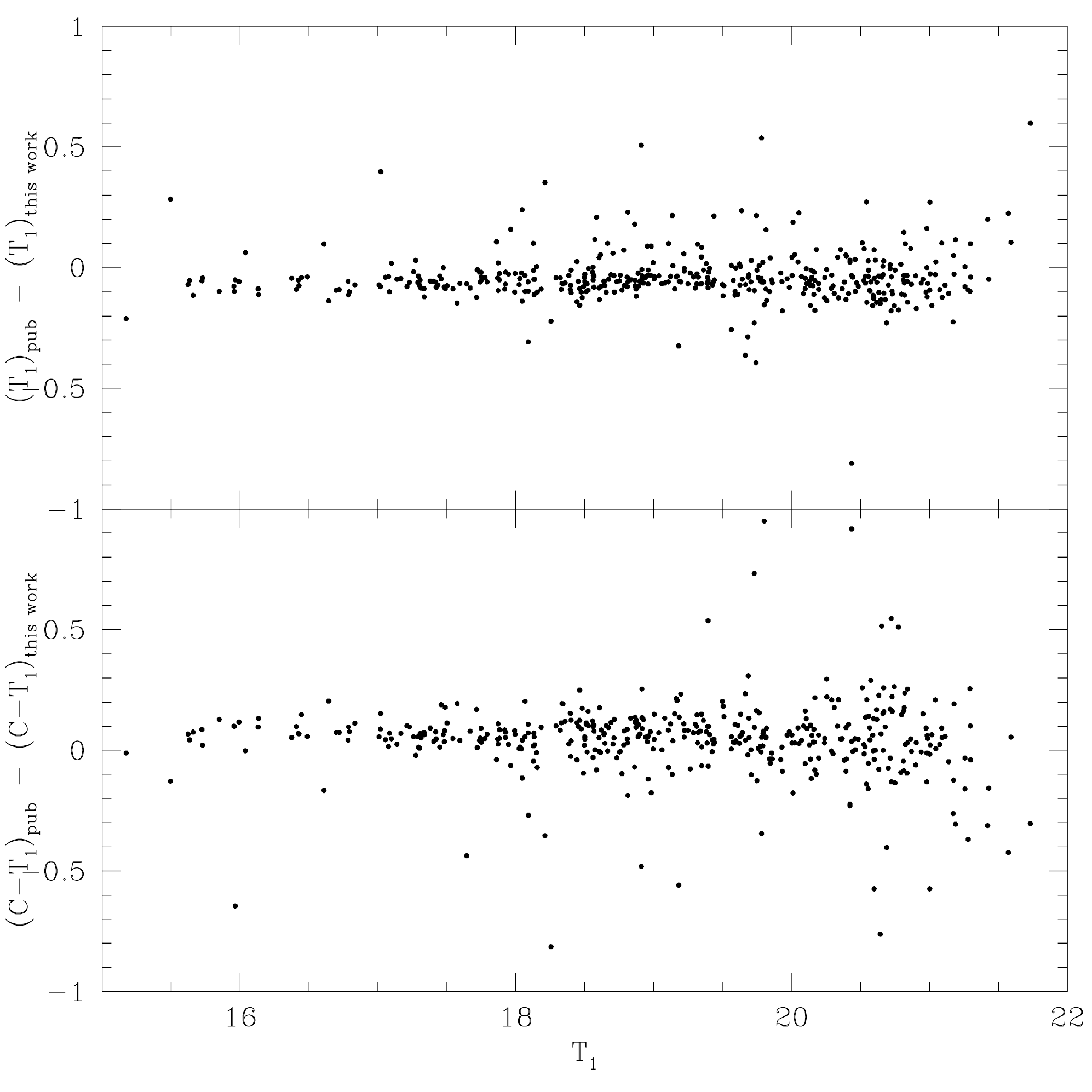}
    \caption{Comparison of $CT_1$ photometry between that of \citet{petal07b} and this
work for the control cluster Lindsay\,106.}
    \label{fig:fig2}
\end{figure}

The calculated internal accuracy of the photometry has been computed according to the criteria given by 
\citet{setal90} and includes random noise, errors in the modelling and centering of the stellar
profile. These internal standard errors provided by DAOPHOT for the $T_1$ magnitude and $C-T_1$ colour have 
been represented by errorbars at the left-hand margin of the observed CMD  
in Fig.~\ref{fig:fig1}.
Fig.~\ref{fig:fig2}  shows how the differences between our $T_1$ magnitudes and $C-T_1$ colours and those obtained by 
\citet{petal07b} vary as a function of $T_1$.  Offsets of 
$\Delta$$T_1$ = $T_{1_{\rm pub}}$ - $T_{1_{\rm this work}}$ = -0.055 $\pm$ 0.092 and 
$\Delta$$(C-T_1)$ = $(C-T_1)_{\rm pub}$ - $(C-T_1)_{\rm this work}$ = 0.050 $\pm$ 0.112
have been computed from 425 stars measured in common.

\begin{table*}
\caption{$CT_1$ data of stars in the field of Haffner\,9.}
\label{tab:table2}
\begin{tabular}{@{}lcccccccc}\hline
Star & $x$ & $y$ & $T_1$ & $\sigma$($T_1$) & n$_{T_1}$ & $C-T1$ & $\sigma$($C-T_1$) & n$_{C-T_1}$  \\
     & (pixel) & (pixel) & (mag)   & (mag)  &  & (mag)    & (mag)  & \\\hline
 -- & --& --& -- & --& --& -- & --& -- \\
   102& 1613.207 & 182.751 &  14.554  &  0.015 & 4 &   1.228  &  0.059 & 4\\
   103&  608.814 & 186.716 &  16.886  &  0.019 & 4 &   1.739  &  0.045 & 4\\
   104& 1012.074 & 193.898 &  15.197  &  0.014 & 4 &   1.350  &  0.029 & 4\\
 -- & --& --& -- & --& --& -- & --& -- \\
\hline
\end{tabular}
\end{table*}

\subsection{$BVI$ photometric data sets}

\citet[][hereafter C13]{carraroetal2013} have carried out a photometric $BVI$ 
campaign focused on
the study of open clusters preferentially located towards the Third Galactic Quadrant.
They used the Y4KCAM camera attached to the  CTIO 1-m telescope, operated by the SMARTS consortium during an observation run in 2005 November-December. In the aforementioned
work they did not include Haffner\,9, although they made their $BVI$ 
photometry publicly available\footnote{VizieR On-line Data Catalog: J/MNRAS/428/502}.

The camera used to obtain those data was equipped with an STA 4064$\times$4064 CCD 
with 15-$\mu$m pixels, yielding a scale of 0.289$\arcsec$/pixel and a
field-of-view (FOV) of 20$\arcmin$$\times$20$\arcmin$. The CCD was operated 
without binning, at a nominal gain of 1.44 e-/ADU, implying a readout
noise of 7 e- per quadrant (four amplifiers). Their available photometry add 
significant value
to our study, because of the larger FOV and the fainter magnitude limit reached
using a different telescope/photometric system setup. We refer the reader to the 
work by
C13 for details concerning the data processing, the measurement
of photometric magnitudes and the standardization of their photometry to the $BVI$
system. 

We compared \citet{carraroetal2013}'s photometry with that obtained by
\citet[][hereafter H08]{hasegawa08} and concluded that the former needs to be
corrected by :

\begin{equation}
V_{\rm std} = V_{\rm C13} -0.9
\end{equation}
 \begin{equation}
B_{\rm std} = (B-V)_{\rm C13}*0.74 + 0.4 
\end{equation}
\begin{equation}
I_{\rm std} = -(V-I)_{\rm C13}*0.87 +Vcarr*0.05 + 0.45 
\end{equation}

\noindent in order to match the latter.
As for the H08 photometry, we used the ridgelines in their CMDs 
-thought to be the mean locus of the fiducial 
cluster sequence-, since the data are not available from the authors.
 
We finally merged our Washington $CT_1$ and the corrected available $BVI$ data sets
into only one master table, which we use in our subsequent an analysis.

\section{Analysis of the photometric data}

The pipeline followed in the analysis of the present photometric data sets involved: 
i) define the cluster central coordinates and  trace its stellar density radial profile;
ii)  derive the cluster structural parameters;
iii) decontaminate the cluster CMD from field stars and; iv) estimate the cluster
fundamental parameters.

\subsection{Cluster extension}

In order to obtain the stellar density radial  profile of Haffner\,9, we started by estimating its geometrical centre.
We did that by fitting Gaussian 
distributions to the star counts in the $x$ and $y$ directions. 
The number of stars projected along the $x$ and $y$ directions were counted within intervals 
of 20, 40, 60, 80, 100 pixel wide, and the Gaussian fits repeated each time. 
The fits of the Gaussians were 
performed using the {\sc ngaussfit} routine in the {\sc stsdas/iraf} package. 
We adopted a single Gaussian and fixed the constant 
to the corresponding background levels (i.e. stellar field density assumed to be uniform) and the linear 
terms to zero. The centre of the Gaussian, its amplitude, and its $FWHM$ acted as variables. 
Finally,
we averaged the five different Gaussian centres with a typical standard deviation
of $\pm$ 40 pixels ($\pm$ 16.0$\arcsec$). 

\begin{figure}
	\includegraphics[width=\columnwidth]{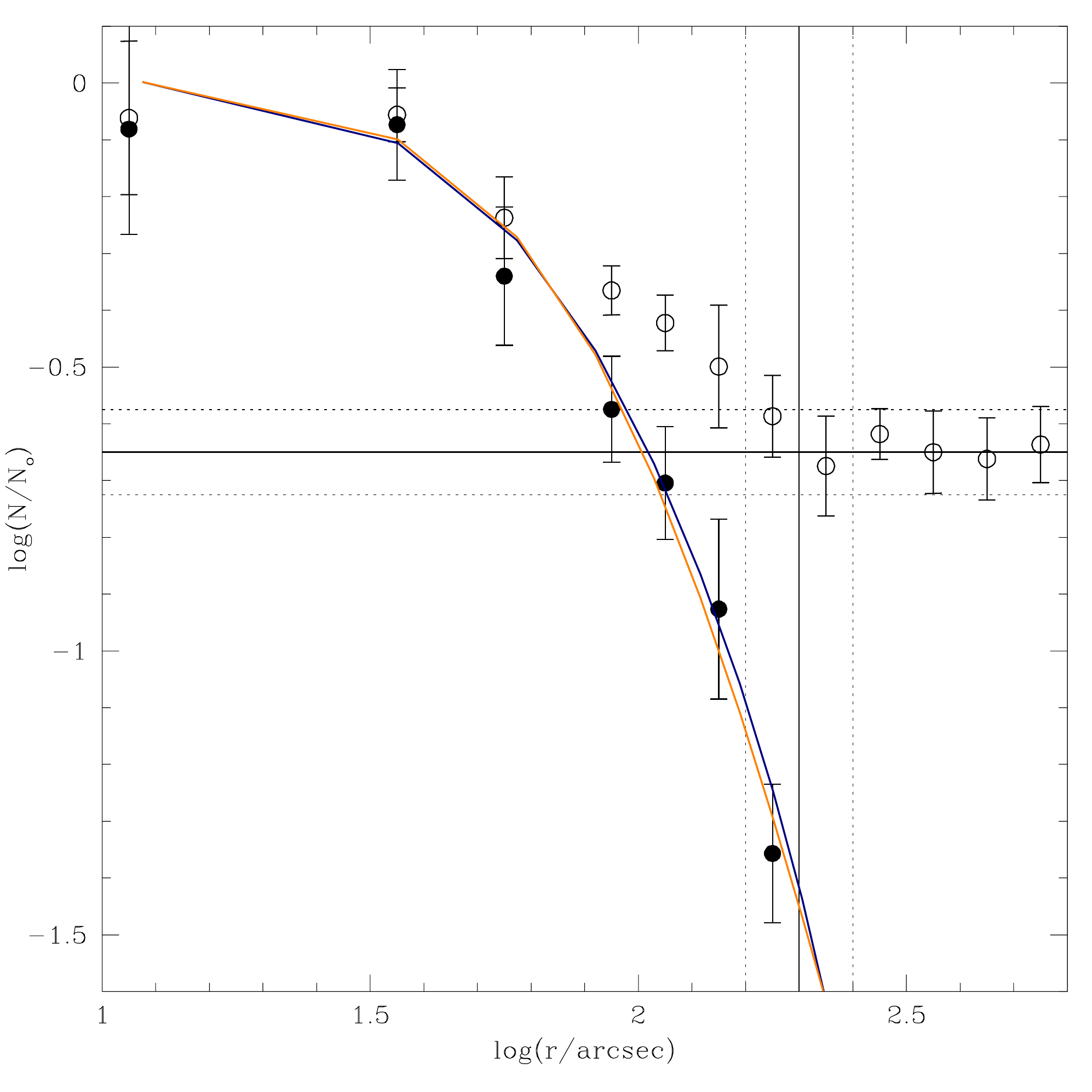}
    \caption{Stellar density profile obtained from star counts. Open and filled circles 
refer to measured and
background subtracted density profiles, respectively. Blue and orange solid lines
depict the fitted King and Plummer curves, respectively.}
    \label{fig:fig3}
\end{figure}

The estimated geometrical centre was then used to built the cluster stellar density profile
from star counts performed within boxes of 60 pixels per side distributed 
throughout the whole observed field.  The chosen box size allowed us to statistically sample  
the stellar spatial distribution.
Thus, the number of stars per unit area at a given radius $r$ can be directly calculated through 
the expression:

\begin{equation}
(n_{r+30} - n_{r-30})/(m_{r+30} - m_{r-30}),
\end{equation}

\noindent where $n_r$ and $m_r$ represent the number of stars and boxes, respectively,  included in a circle of radius $r$. 
The advantage of this method over the frequent counting of stars in annular regions around the cluster centre relies on 
the fact that is not required a complete circle of radius $r$ within the observed 
field to compute the mean stellar density at that distance. Therefore, it is possible to estimate the background level with 
high precision using regions located far away from the cluster centre. With a good placement of the background level,
the cluster radius ($r_{cls}$) results, in turn, in a more reliable estimate. Fig.~\ref{fig:fig3} shows with
open and filled circles the observed and background subtracted density profiles -expressed as number of stars per 
arcsec$^2$-, respectively, while the errobars represent the rms
errors. In the case of the background subtracted density profile we added the mean error of the
background star counts. The background level and the cluster radius ($r_{cls}$ = 200$^{\rm +50}_{\rm -40}$ arcsec) 
are indicated by solid horizontal
and vertical lines, respectively; their uncertainties are in dotted lines.

We fitted the background corrected density profile using both \citet{king62} and
\citet{plummer11} models in order to get independent estimates of the cluster core ($r_c$), half-mass ($r_h$)
and tidal ($r_t$) radii, respectively. We used a grid of $r_c$, $r_h$ and $r_t$ values spanning the
known range of radii of open clusters \citep{piskunovetal2007} and minimised $\chi$$^2$. 
We finally derived  $r_c$ = 70 $\pm$ 10 arcsec, $r_h$ = 125 $\pm$ 15 arcsec and $r_t$ = 400 $\pm$ 100 arcsec,
respectively. In Fig.~\ref{fig:fig3} we superimposed the respective King's and Plummer's curves
with blue and oorange solid lines, respectively.

\subsection{Field star decontamination}

Because Fig.~\ref{fig:fig1} reveals that both cluster and field star sequences are 
more or less superimposed, 
we must firstly separate the cluster stars from those belonging to the surrounding fields on 
a statistical
basis in order to meaningfully use that CMD to estimate the cluster parameters. Note that both cluster and 
field stars are affected by nearly the same interstellar reddening, which is indeed 
what causes the overlapping of their main sequences (MSs). This fact makes the cleaning of the cluster
CMD even more challenging.

To filter the field stars from the CMDs, we applied a statistical procedure which consists, 
firstly, in adopting two CMDs from different regions located far from the cluster. The dimension of 
each selected field region was $\pi$$r_{cls}$$^2$ and acted as references to statistically filter 
an equal circular area centred on Haffner\,9. 

Secondly, by starting with reasonably large boxes -- typically 
($\Delta$($T_1$),$\Delta$($C-T_1$)) = (1.00, 0.50) mag -- centred on each
star in both field CMDs and by subsequently reducing their sizes
until they reach the stars closest to the boxes' centres in magnitude and colour,
separately, we defined boxes which result in use of larger areas in
field CMD regions containing a small number of stars, and vice versa.
Note that the definition of the position and size of each box involves
two field stars, one at the centre of the box and another -the closest one to 
box centre - placed on the boundary of that box. \citet{pb12} have shown that this is
an effective way of accounting for the local field-star
signature in terms of stellar density, luminosity function and/or
colour distribution.

\begin{figure*}
	\includegraphics[width=\textwidth]{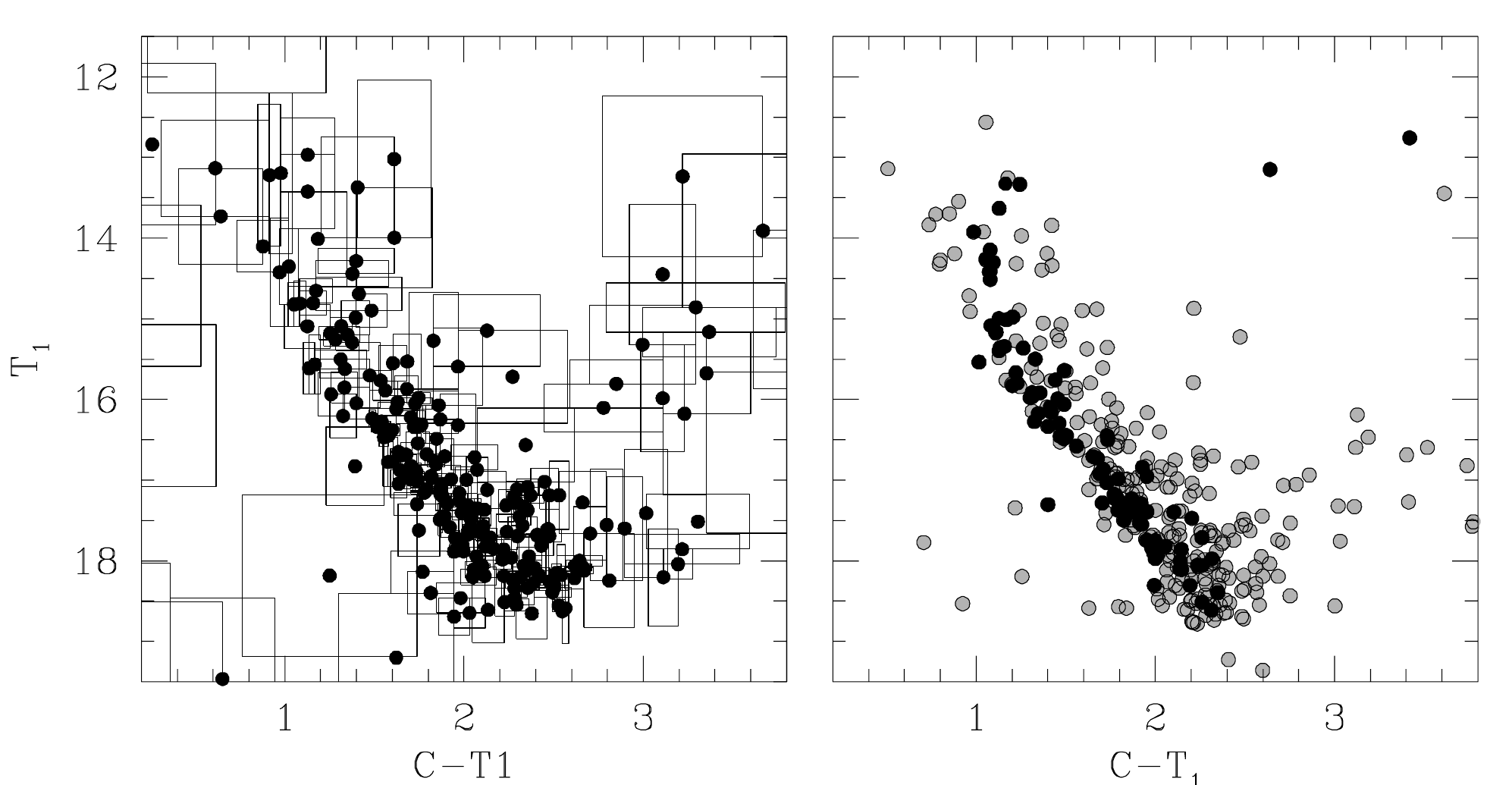}
    \caption{CMDs for stars in the field of Haffner\,9: a field CMD for annular region centred 
on the cluster and with a size equal to the cluster area, and the corresponding
defined set of boxes overplotted (left-hand panel) and; the observed and cleaned CMD
  composed of the stars distributed within the cluster radius represented by gray and black
filled circles, respectively  (left-hand panel).}
    \label{fig:fig4}
\end{figure*}

Next, we plotted all these boxes
for each field CMD on the cluster CMD and subtracted the star located
closest to each box centre. Since we repeated this task for each of the two 
field CMD box samples, we found that some stars have remained unsubtracted once
or twice. Finally, we adopted as the cluster CMD that built from stars
that have not been subtracted any time. In order to illustrate the performance of
the statistical cleaning procedure, Fig.~\ref{fig:fig4} depicts a single field-star CMD 
(left-hand panel) for an annular region around Haffner\,9 with an area equal to that of the 
cluster, and the cleaned cluster CMD superimposed to the observed ones represented by black and gray
filled circles, respectively (right-hand panel). As can be seen, differences in stellar composition
become noticeable when comparing field and cleaned cluster CMDs. Particularly, the cluster
evolved MS is now clearly seen.

\subsection{Cluster's fundamental parameters}

\begin{figure*}
	\includegraphics[width=\textwidth]{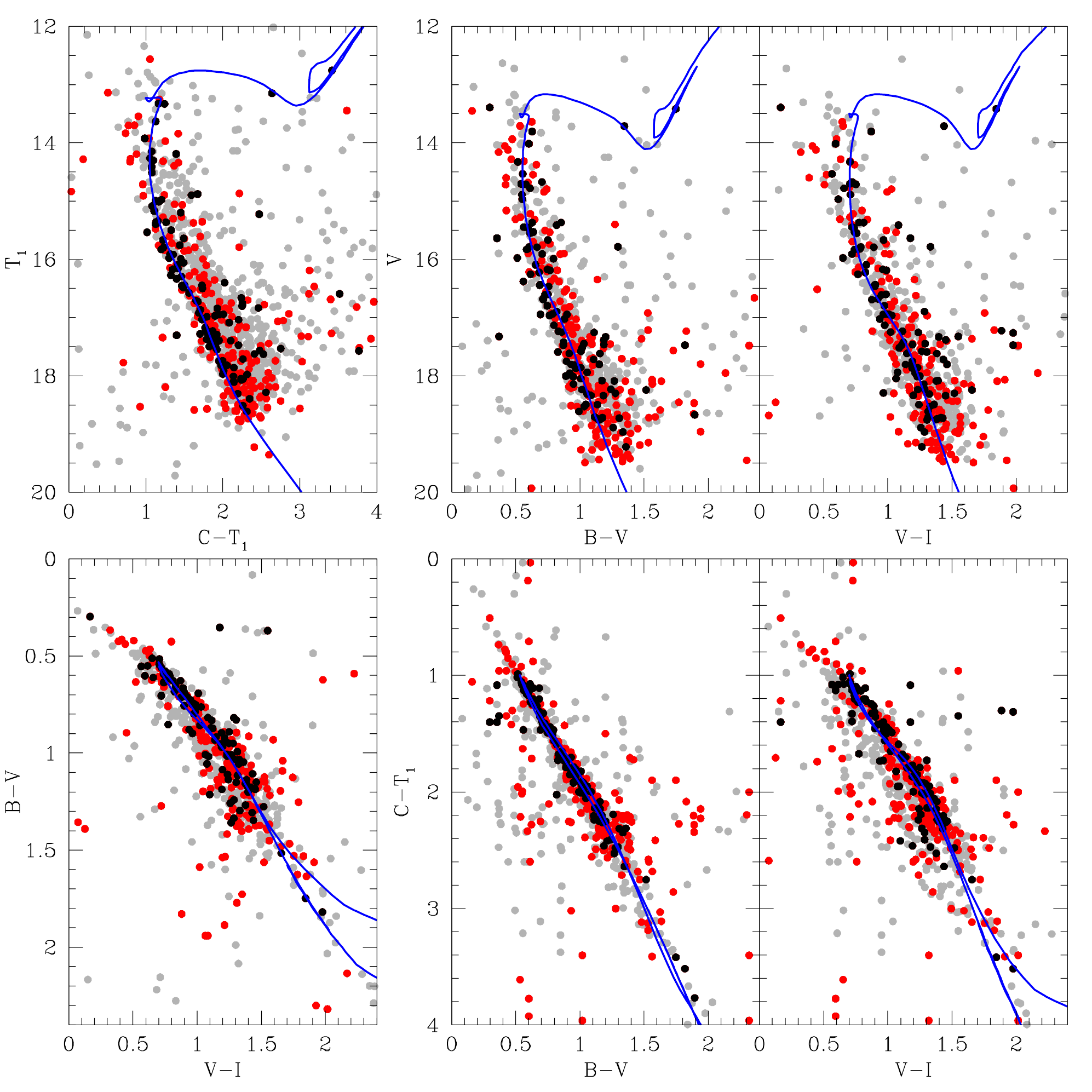}
    \caption{CMDs and CC diagrams for stars measured in the field of Haffner\,9. 
Grey, red and black filled circles represent all the measured stars, those within
the cluster radius and the cluster stars from the field star cleaning
procedure, respectively. We overplotted the isochrone which best matches the
cluster features (see text for details).}
    \label{fig:fig5}
\end{figure*}

We present in Fig.~\ref{fig:fig5} the whole set of CMDs and colour-colour (CC) diagrams
for Haffner\,9 that can be exploited from the present multi-band photometry.
They include every magnitude and colour for the whole sample of observed stars, those located
within the cluster radius and those considered cluster stars from the field star decontamination
procedure plotted with grey, red and black filled circles, respectively.
At first glance, the cleaned
cluster CMDs resemble those of a cluster with a moderate age, projected on to
a star field not easy to disentangle from the cluster.

The availability of three CMDs and three different CC diagrams covering wavelengths from
the blue up to the near-infrarred
allowed us to derive reliable ages, reddenings and distances for Haffner\,9 from the
matching of theoretical isochrones. 
In order to enter the isochrones into the CMDs and CC diagrams we used the following
ratios: $E(V-I)$/$E(B-V)$ = 1.25, $A_{V}$/$E(B-V)$ = 3.1 \citep{cetal89}; 
$E(C-T_1)$/$E(B-V)$ = 1.97 and $A_{T_1}$/$E(B-V)$ = 2.62 \citep{g96}.

We started by selecting theoretical isochrones \citep{betal12} with  solar metal content
in order to choose that which best match the cluster`s features in the CMDs. 
In this sense, the shape of the MS, its curvature, the relative 
distance between the red giants and the MS turn-off (MSTO) in magnitude and colour 
separately, among others, are features tightly related to the cluster age, regardless 
their reddenings and distances. Note also that, by considering the whole metallicity 
range of the Milky Way open
clusters \citep[see, e.g.][]{paunzeretal2010,hetal14} and by using the theoretical isochrones of 
\citet{betal12},
the differences at the zero age main sequence (ZAMS) in $V-I$ colour is
smaller than $\sim$ 0.08 mag. This result implies that
negligible differences between the ZAMSs for the cluster metallicity and that of solar
metal content would appear, keeping in mind the intrinsic 
spread of the stars in the $V$ vs $V-I$ CMD. 

From our first
choices, we derived the cluster reddening by shifting those isochrones in the three CC diagrams
following the reddening vectors until their bluest points coincided with the observed ones.
Note that this requirement allowed us to use the three CC diagrams, even though the reddening vectors run almost parallell to the cluster sequence.
Finally, the mean $E(B-V)$ colour excesse was used to properly shift the chosen isochrones
in the three CMDs in order to derive the cluster true distance modulus by shifting the isochrones 
along the magnitude axes. We iterated this procedure for different ages as well as for
metallicities [Fe/H] = -0.5 up to 0.2 dex. We found that isochrones bracketing the cluster 
age (log($t$ yr$^{-1}$) = 8.55) by $\Delta$log($t$ yr$^{-1}$) = $\pm$0.05 and the cluster metallicity ([Fe/H] = 0.0 dex) by 
$\Delta$[Fe/H] = $\pm$0.1 dex represent the overall age and metallicity uncertainties
owing to the observed dispersion in the cluster CMDs and CC diagrams. 
As for the cluster reddening and true distance modulus, we obtained $E(B-V)$ = 0.60 $\pm$ 0.05
mag and $m-M_o$ = 12.5 $\pm$ 0.2 mag, respectively. Fig.~\ref{fig:fig5} shows
the adopted best matched isochrone overplotted on to the CMDs and CC diagrams with a blue
solid line.

\begin{figure}
	\includegraphics[width=\columnwidth]{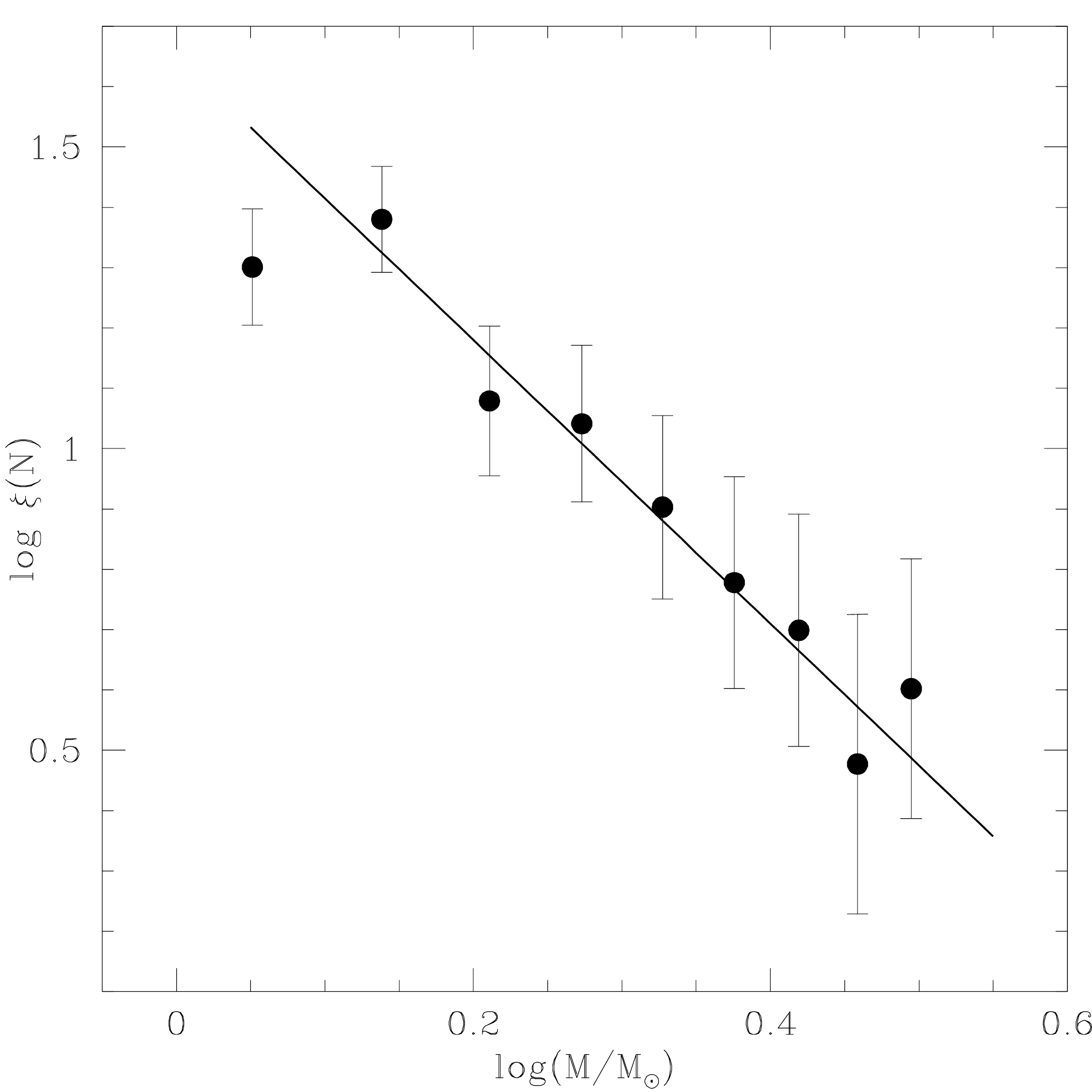}
    \caption{The present mass function of Haffner\,9. The \citet{salpeter55}' relationship for 
stars in the solar neighbourhood is superimposed with a solid line.}
    \label{fig:fig6}
\end{figure}

The cluster mass and its present mass function (MF) were derived by summing the 
individual masses of stars not eliminated during the complete cleaning procedure. 
Those individual masses were obtained by interpolation 
in the theoretical isochrone traced in Fig.~\ref{fig:fig5} from the 
observed $T_1$ magnitude of each star, properly corrected by reddening and distance modulus.
We finally attained $\log(M_{cls}/\msun)$ = 2.2 $\pm$ 0.2. The uncertainty comes from propagation of the $T_1$ magnitude errors in the
mass distribution along the theoretical isochrone as well as from considering stars
subtracted once in the field star cleaning procedure. Note that, when building the 
stellar density profile from stars not eliminated at any time, we obtained a 
curve which matches the 
King's curve drawn in Fig.~\ref{fig:fig3}. This very good agreement implies that 
the cleaning procedure subtracted an appropriate number of stars according to the stellar density of the backgroung/foreground field, so that the mass uncertainty would be
smaller if we did not considered stars subtracted once.
The resulting MF is
shown in Fig.~\ref{fig:fig6} where the errorbars come from applying Poisson statistics. 
For comparison porpuses we superimposed the relationship
given by \citet[][slope = -2.35]{salpeter55} for the stars in the solar neighbourhood.

Using the resulting mass and the half-mass radius $r_h$, we
computed the half-mass relaxation times using the equation \citep{sh71}:

\begin{equation}
t_r = \frac{8.9\times 10^5 M_{cls}^{1/2} r_h^{3/2}}{\bar{m} log_{10}(0.4M_{cls}/\bar{m})}
,\end{equation}

\noindent where $M_{cls}$ is the cluster mass and $\bar{m}$ is the mean mass of 
the cluster stars ($\bar{m}$ = 1.7 \msun). We obtained $t_r$ = 16 $\pm$ 2 Myr. 
If we considered non-oberved stars with masses between 1 and 0.5 $\msun$ and the 
Salpeter's mass function, the relaxation times would increase in $\sim$ 10 per cent.
Note that, despite the advanced state of dynamical evolution of Haffner\,9 
($<$age/$t_r$$>$ = 22), the cluster still keeps its MF close to that of Salpeter's law
(see Fig.~\ref{fig:fig6}).

Finally, we employed the \texttt{ASteCA} 
suit of functions \citep{perrenetal2015} to generate $\approx$ 2.2$\times$10$^5$ synthetic CMDs
of a star cluster covering ages from log($t$ yr$^{-1}$) = 8.5 up to 8.6 
($\Delta$log($t$ yr$^{-1}$) = 0.01),  metallicities in the range $Z$ = 0.012 - 0.019 
($\Delta$$Z$ = 0.001), interstellar extinction between 0.55 and 0.65 mag 
($\Delta$$E(B-V)$ = 0.01 mag), distance modulus between 12.3 and 12.7 mag  
($\Delta$$(m-M)_o$ = 0.05 mag) and total mass in the range 100 - 250 $\msun$ 
($\Delta$$M$ = 10$\msun$), respectively. 

The steps by which a synthetic star cluster for a given set of age, metalicity, distance modulus, and reddening values is generated by \texttt{ASteCA} is as follows: i) a theoretical isochrone
is picked up, densely interpolated to contain a thousand points throughout its entire length,
including the most evolved stellar phases. ii) The isochrone is shifted in colour and
magnitude according to the $E(B-V)$ and $(m-M)_o$ values to emulate the effects these extrinsic 
parameters have over the isochrone in the CMD. 
iii) The isochrone is trimmed down to a certain faintest magnitude 
according to the limiting magnitude thought to be reached. iv) An initial mass function 
(IMF) is sampled in the mass range $[{\sim}0.01{-}100]\,M_{\odot}$ up
to a total mass value $M$ provided that 
evolved CMD regions result properly populated.
The distribution of masses is then used to obtain a properly populated synthetic 
star cluster by keeping one star in the interpolated 
isochrone for each mass value in the distribution. v) A random fraction of stars are 
assumed to be binaries, which is set by default to  
$50\%$ \citep{von_Hippel_2005}, with secondary masses 
drawn from a uniform distribution between the mass of the primary star and a 
fraction of it given by a mass ratio parameter set to $0.7$. 
vi) An appropriate
magnitude completeness and an exponential photometric error functions are
finally applied to the synthetic star cluster. 

\begin{figure}
	\includegraphics[width=\columnwidth]{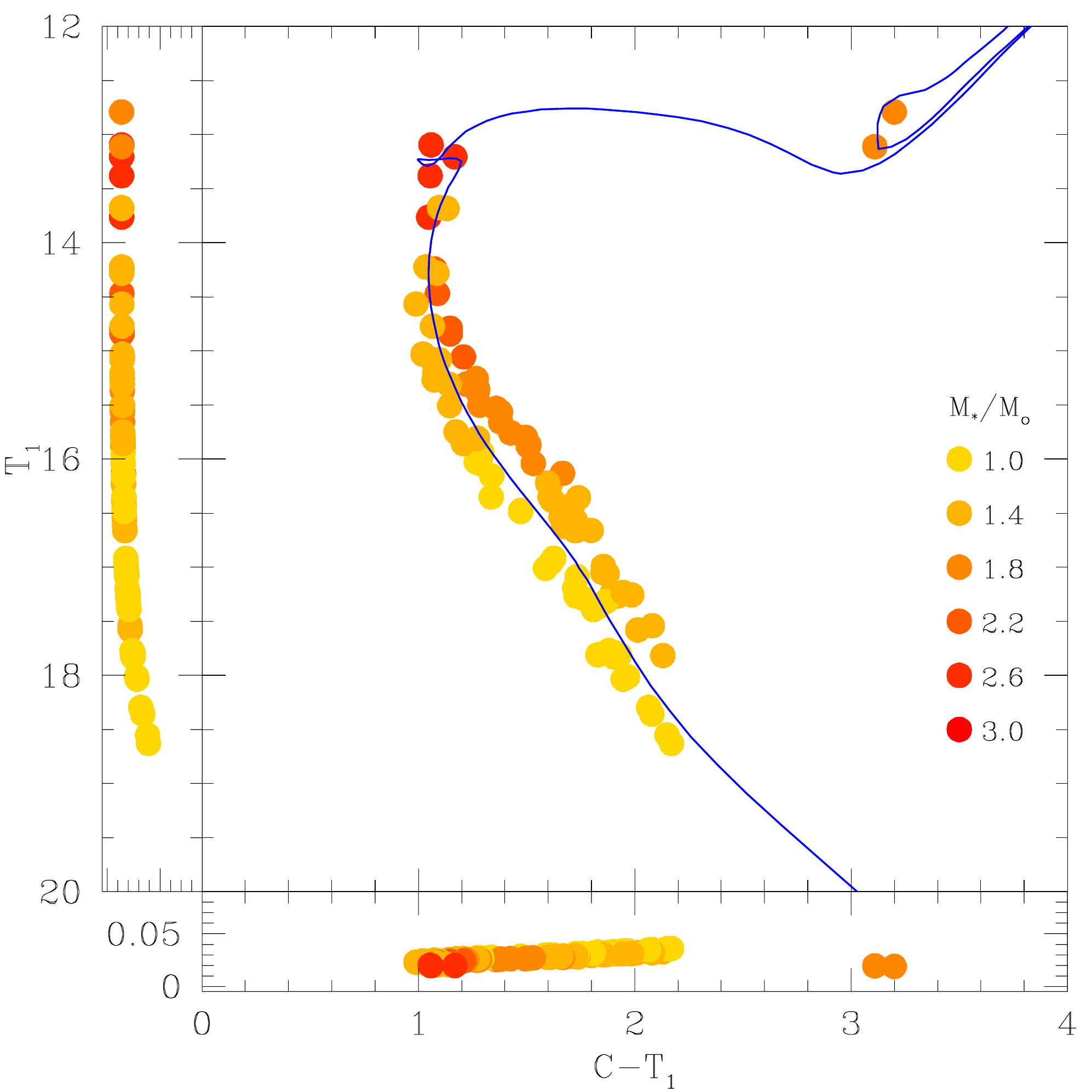}
    \caption{Best-generated star cluster CMD with the uncertainties in $T_1$ and
$C-T_1$, the stellar masses in colour-scaled filled circles, and the 
theoretical isochrone for log($t$ yr$^{-1}$) = 8.55 and [Fe/H] = 0.0 dex superimposed.}
    \label{fig:fig7}
\end{figure}

Fig.~\ref{fig:fig7} shows the synthetic CMD which best matches the cluster's parameters, 
with the generated uncertainties in $T_1$ and $C-T_1$, the range of stellar masses drawn in
colour-scaled filled circles and the theoretical isochrone for log($t$ yr$^{-1}$) =
8.55 and [Fe/H] = 0.0 dex superimposed.

\section{Analysis and Discussion}

\begin{figure*}
	\includegraphics[width=\textwidth]{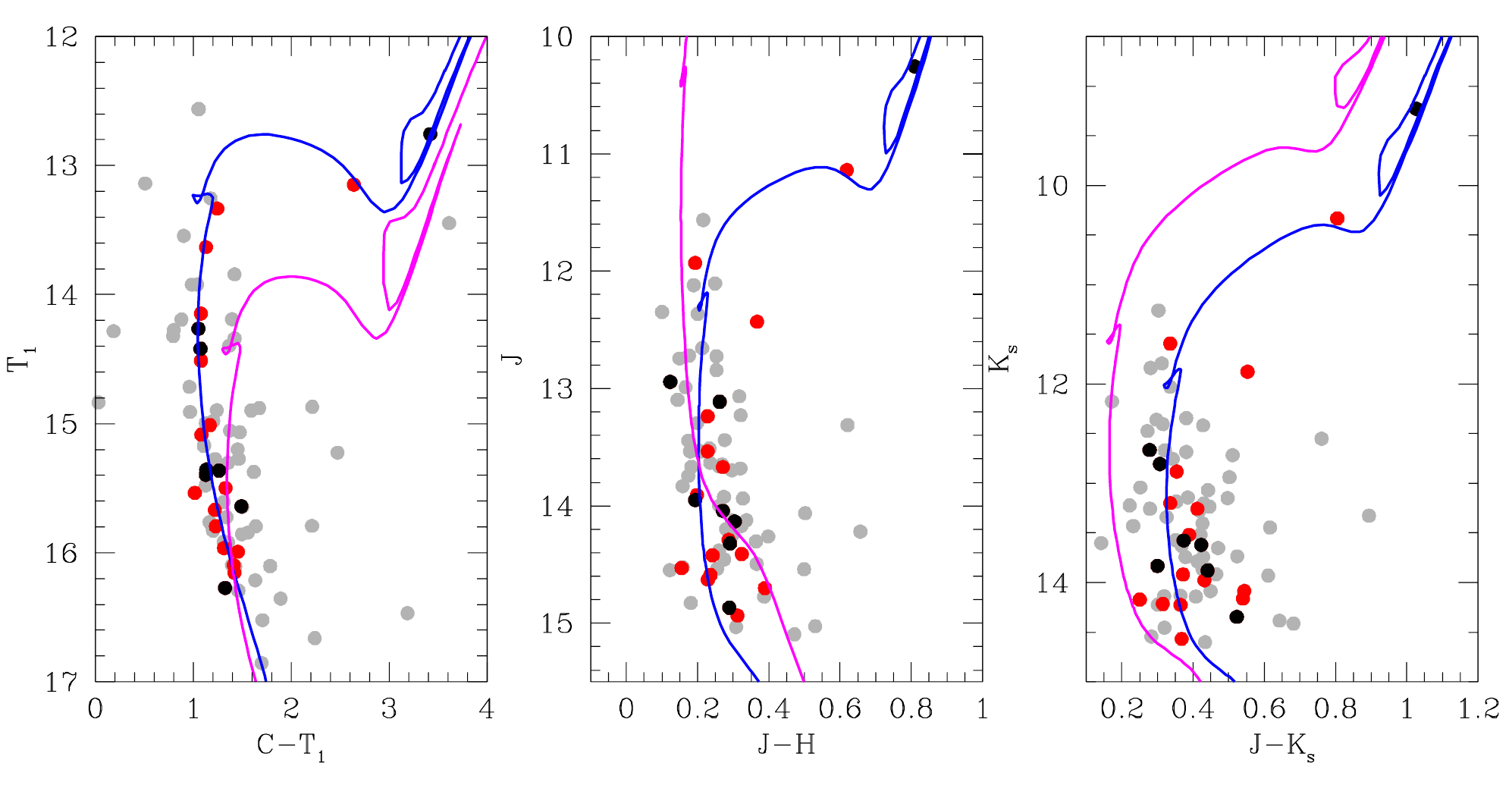}
    \caption{CMDs for stars in the field of Haffner\,9 with proper motion measurements
(grey filled circles). Stars with proper motion membership probabilities higher than
75 and 90 per cent are drawn with red and black filled circles, respectively. The
isochrones adopted in this work is superimposed with a blue solid line, while those
for the parameters estimated by \citet{hasegawa08}, \citet{bb05} and \citet{bf14} are
plotted with a magenta line in the left-hand, middle and right-hand panels, respectively
(see text in Section 4 for details.}
    \label{fig:fig8}
\end{figure*}

\citet{diaswetal2014} presented a catalog of mean proper motions and membership probabilities 
of individual stars for optically visible open clusters, among them, Haffner\,9,
 using data from the UCAC4 catalog \citep{zachariasetal2013}. We cross-correlated
the stars with proper motions in our data sets and built the corresponding $T_1$ vs. 
$C-T_1$ CMD. The left-hand panel of Fig.~\ref{fig:fig8} shows with grey, red and black
filled circles all these stars, those with membership probabilities $P$ $>$ 75 and 90 per cent, respectively. As can be seen, the proper motion memberships confirm that the cleaned
CMD produced in Section 3.2, which is very well matched by an isochrone of log($t$ yr$^{-1}$) = 8.55 
and [Fe/H] = 0.0 dex (see details in Section 3.3), corresponds to that of the cluster.
This is an important probe, since previous detailed photometric studies of this object
led to different fundamental parameters.

For instance, \citet{bb05} and \citet{kharchenko13} used 2MASS data \citep{skrutskieetal97},
separately, and derived an age of 140 $\pm$ 20 Myr, a true distance modulus  $(m-M)_o$ =
11.40 $\pm$ 0.1 mag, and a reddening $E(B-V)$ =  0.50 $\pm$ 0.05 from the fit of
theoretical isochrones in the $J$ versus $J-H$ CMD. However, \citet{bf14}, also from
2MASS data, obtained an age of 250 $\pm$ 100 Myr, $(m-M)_o$ =
12.40 $\pm$ 0.4 mag, and $E(B-V)$ =  0.40 $\pm$ 0.10, respectively. All three studies
assumed a solar metal content. In order to seek 
for any source of discrepancy in these 2MASS data analyses, we took advantage of the
photometric and proper motion membership probabilities analyzed above. Fig.~\ref{fig:fig8}
(middle and right-hand panels) depict the 2MASS CMDs used by \citet{bb05} and \citet{bf14}
to derive the cluster's fundamental parameters. We plotted every star with proper motion using
the same colour code as in the left-hand panel. Note that the fainter magnitude limit
reached by stars with proper motion measurements is nearly similar to that of the 2MASS data.
For the sake of the reader, we superimposed
the isochrone of log($t$ yr$^{-1}$) = 8.55 and [Fe/H] = 0.0 dex as well as those for the 
reddenings, distance modulii, ages and metallicities used by \citet{bb05} and \citet{bf14}
with  blue and magenta solid lines, respectively.

From the examination of these CMDs some conclusions can be drawn. Firstly, the magnitude
depth of 2MASS data is noticeable much brigher than that of the present data set (see
Fig.~\ref{fig:fig5}), whereas the colour baseline of the infrared colours is much shorter
than that of $C-T_1$, which do not favour an accurate isochrone matching. Secondly,
the isochrone adopted in this work reasonably matches the cluster sequences, although the
scatter is larger than that seen in the $T_1$ vs. $C-T_1$ CMD. Seemingly, \citet{bb05}
have fitted a MS with some field contamination (middle panel). In the case of \citet{bf14}
we especulate with the possibility that they could have considered a larger cluster area
with field stars with bluer infrared colours and no proper motions. 

\citet{hasegawa08} presented $BVI$ photometry of stars in the cluster field and
estimated an age of 500 Myr, a metal content [Fe/H] = -0.4 dex,
a true distance modulus of $(m-M)_o$ = 12.7 mag, and a reddening $E(B-V)$ =  0.75 mag.
For comparison purposes, we plotted such an isochrone in the left-hand panel of
Fig.~\ref{fig:fig8} with a magenta solid line. Once again, the employment of
a field star contaminated CMD could lead them to derive unreliable cluster parameters.

\begin{figure}
	\includegraphics[width=\columnwidth]{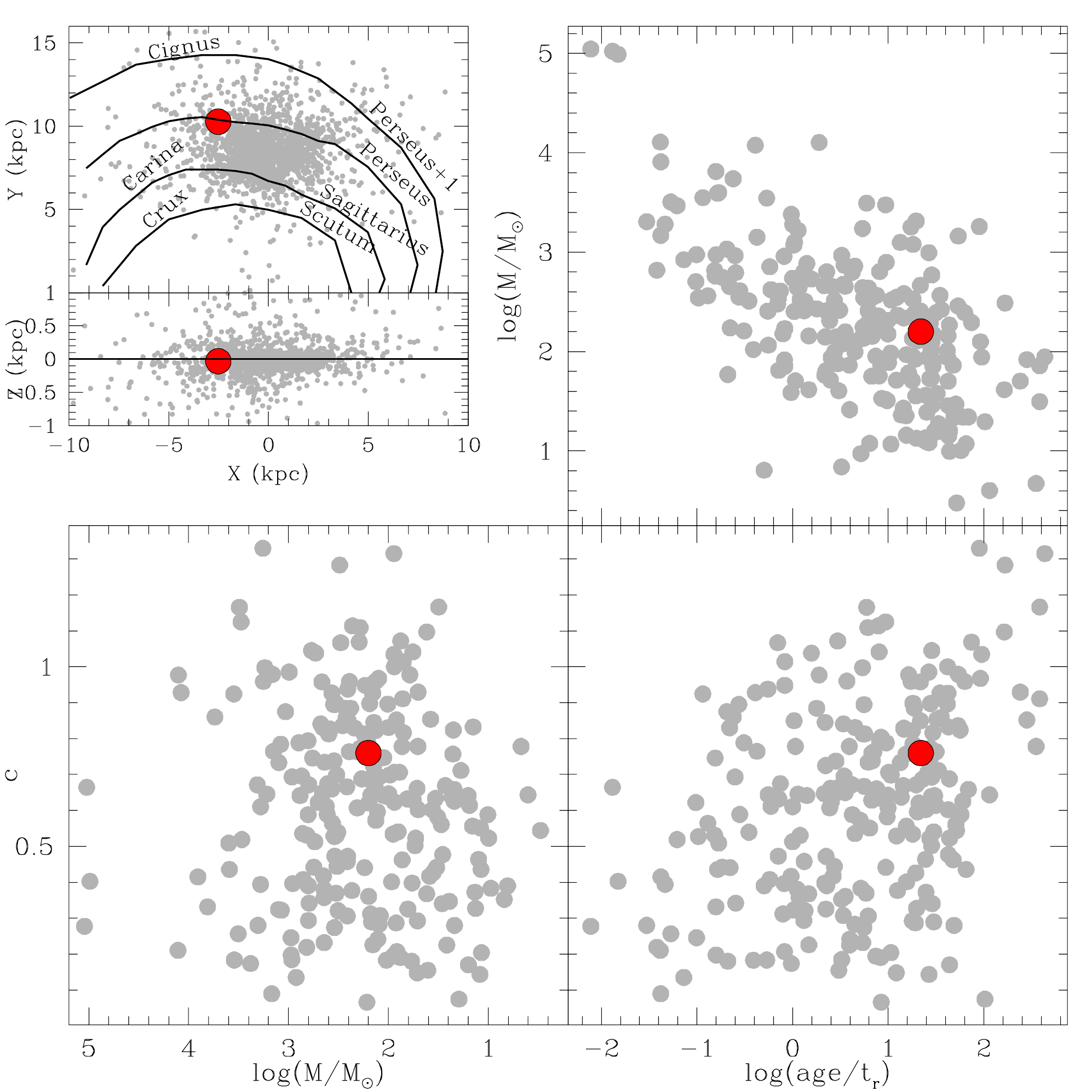}
    \caption{{\it Top-left}: Galactic spatial position of Haffner\,9 (red filled cirled). 
Open clusters from the
catalogue of \citet[][version 3.5 as of January 2016]{detal02} are drawn with gray dots, while
the schematic positions of spiral arms \citep{ds2001,moitinhoetal2006} are traced with black 
solid lines. {\it Top-right and bottom}:
Relationships between cluster concentration parameter ($c$), mass, age and relaxation time ($t_r$). Grey dots correspond to 236 star clusters with homogeneous 
estimations of masses and radii derived by \citet{piskunovetal2007}.
}
    \label{fig:fig9}
\end{figure}

We computed Galactic coordinates using the derived cluster heliocentric distances, 
their angular
Galactic coordinates and a Galactocentric distance of the Sun of R$_{GC_\odot}$ = 8.3 kpc 
\citep[][and references therein]{hh2014}. The resulting spatial distribution is depicted
in the top-left panel of
Fig.~\ref{fig:fig9}, where we added for comparison purposes the 2167 open clusters 
catalogued by 
\citet[][version 3.5 as of January 2016]{detal02} and the schematic positions of the spiral
arms \citep{ds2001,moitinhoetal2006}. Haffner\,9 is
located at the Perseus arm, beyond  the cirle around the Sun 
(d $\sim$ 2.0 kpc) where the catalogued clusters are mostly concentrated.

The age/$t_r$ ratio is a
good indicator of the internal dynamical evolution, since it gives the number of
times the characteristic time-scale to reach some level of energy equipartition 
\citep{bm98} has been surpassed. Star clusters with large age/$t_r$ ratios have reached a
higher degree of relaxation and hence are dynamically more evolved. As 
Fig.~\ref{fig:fig9} shows, Haffner\,9 appears to have had enough time to evolve dynamically. In the figure we included in grey colour 236 
open clusters analysed by \citet{piskunovetal2007}, who derived from them homogeneous
scales of radii and masses. They derived core and tidal radii for their cluster sample,
from which we calculated the half-mass radii and, with their clusters masses and eq. 4, relaxation times, by assuming that the cluster stellar density
profiles can be indistinguishably reproduced by King and Plummer models.
Their cluster sample are mostly distributed inside 
a circle of $\sim$ 1 kpc from the Sun. As compared to the \citet{piskunovetal2007}'s sample,
Haffner\,9 is placed towards the most
evolved limit of the age/$t_r$ distribution (right-hand panels).

Since dynamical evolution implies the loss of stars (mass loss), we expect some trend of
the present-day cluster mass with the age/$t_r$ ratio. This is confirmed
in the top-right panel of Fig.~\ref{fig:fig5}, where the larger the present-day
mass the less the dynamical evolution of a cluster in the solar neighbourhood,
with a noticeable scatter. Haffner\,9 appears to have a relatively large mass
for its particular internal dynamical state. Curiously, selection against poor 
and old clusters could suggest the beggining of cluster dissolution, with some exceptions.

\citet{trentietal2010} presented a unified picture for the evolution of star clusters 
on the two-body relaxation timescale from direct N-body simulations of star clusters
in a tidal field. Their treatment of the stellar evolution is based on the approximation
that most of the relevant stellar evolution occurs on a timescale shorter than a 
relaxation time, when the most massive stars lose a significant fraction of mass and consequently  contribute to a global expansion of the system. Later in the
life of a star cluster, two-body relaxation tends to erase the memory of the initial 
density profile and concentration. They found that the
structure of the system, as measured by the core to half mass radius ratio, the
concentration parameter $c$= log($r_t/r_c$), among others, 
evolve toward a universal state, which is set by the efficiency of heating on
the visible population of stars induced by dynamical interactions in the core of
the system. In the bottom panels of Fig.~\ref{fig:fig9} we plotted the
dependence of the concentration parameter $c$ with the cluster mass and the
age/$t_r$ ratio, respectively. They show that our dynamically evolved cluster is 
within those with relatively high $c$ values, and that star clusters tend to 
initially start their dynamical evolution with relatively small concentration 
parameters. Likewise, star clusters in an advanced dynamical state can also have relatively
lower $c$ values due to their smaller masses.

\section{Summary and Conclusions}

In order to continue our Washington $CT_1$ and Johnson $BVI$ photometric studies 
on Milky Way star clusters, 
we turned our attention to Haffner\,9, a previously studied open cluster with
a noticeable spread in the values obtainted of its fundamental parameters.

The analysis of the current photometric data sets leads to the following main conclusions:

(i) To disentagle cluster features from those belonging to their surrounding fields, 
we applied a subtraction procedure to statistically clean the cluster CMDs from field star contamination. The employed 
technique makes use of variable cells in order to reproduce the field CMD as closely 
as possibly. The stellar density profile built from stars that reamined unsubtracted
very well matches that obtained from star counts carried out throughout the observed
field, once the background level is subtracted. Moreover, the main cluster features in the
cleaned CMDs are confirmed when proper motion membership probabilities are taken into
account.

(ii) Using the cleaned cluster CMDs and CC diagrams, we estimated the cluster
fundamental parameters in a self-consistent way. The availability of three CC diagrams 
and three CMDs covering wavelengths from 
the blue up to the near-infrarred allowed us to derive reliable values of
age, metallicity, reddenings and 
distances for Haffner\,9. We exploited such a wealth in combination with
theoretical
isochrones computed by \citet{betal12} to find out that the cluster
is 350 Myr old, is placed in the Perseus arm at a solar heliocentric distance of 3.2 kpc,
and has nearly solar metal content. The lower limits of its present mass is 
$\sim$ 160$\msun$. We confirmed such a limit from the generation of thousand synthetic
CMDs.

(iii) We found that a less deep photometry, a narrower colour baseline and a
less effective CMD cleaning procedure, could have been some sources  
that led previous studies to derive cluster fundamental parameters which do not
correspond to the fiducial cluster features. By using the same photometric data sets
as in previous works and proper motion
memberships, we confirm the present cluster fundamental parameters.

(iv) Finally, we estimated the half-mass relataxion time for Haffner\,9, which turned
out to be $\sim$ 22 smaller than the cluster age. This result suggests that Haffner\,9
is facing an advanced state of its internal dynamical evolution. However, the cluster 
still keeps its MF close to that of the Salpeter's law.
When combined with the obtained structural
parameters, we found that the cluster is possibly in the phase typical of those
with mass segregation in their core regions.

\section*{Acknowledgements}

We thank Giovanni Carraro for revising the manuscript and making useful
suggestions.
We thank the anonymous referee whose thorough comments and suggestions
allowed us to improve the manuscript.

%%%%%%%%%%%%%%%%%%%%%%%%%%%%%%%%%%%%%%%%%%%%%%%%%%

%%%%%%%%%%%%%%%%%%%% REFERENCES %%%%%%%%%%%%%%%%%%

% The best way to enter references is to use BibTeX:

\bibliographystyle{mnras}
%\bibliography{paper} % if your bibtex file is called paper.bib

%to be uncommented before sending to editor
\input{paper.bbl}

% Alternatively you could enter them by hand, like this:
% This method is tedious and prone to error if you have lots of references
%\begin{thebibliography}{99}
%\bibitem[\protect\citeauthoryear{Author}{2012}]{Author2012}
%Author A.~N., 2013, Journal of Improbable Astronomy, 1, 1
%\bibitem[\protect\citeauthoryear{Others}{2013}]{Others2013}
%Others S., 2012, Journal of Interesting Stuff, 17, 198
%\end{thebibliography}

%85 180 592 718    1
%310 180 592 718   2
%85 440 592 718    3
%310 440 592 718   4

%%%%%%%%%%%%%%%%%%%%%%%%%%%%%%%%%%%%%%%%%%%%%%%%%%

%%%%%%%%%%%%%%%%% APPENDICES %%%%%%%%%%%%%%%%%%%%%

%If you want to present additional material which would interrupt the flow of the main paper,
%it can be placed in an Appendix which appears after the list of references.

%%%%%%%%%%%%%%%%%%%%%%%%%%%%%%%%%%%%%%%%%%%%%%%%%%

% Don't change these lines
\bsp	% typesetting comment
\label{lastpage}

% End of mnras_template.tex

\end{document}